\documentclass[twocolumn,usenatbib]{aa} 
\usepackage{graphicx} 
\usepackage{epsfig,graphicx,rotating,portland,fancyheadings,longtable,natbib,amsmath} 
\def\3c{3C~454.3}
\def\sw{Swift~}
\def\sax{{\it Beppo}SAX}
\def\ergs{{erg~cm$^{-2}$s$^{-1}$~}} 
\def\phcm2{{ph$^{-2}$s$^{-1}$~}} 
\begin{document}

\title{\sw and infra-red observations of the blazar 3C~454.3\\  during the giant X-ray flare of May 2005}

\author{P.~Giommi\inst{1}$^,$\inst{2}, A. J.~Blustin\inst{3}, 
M.~Capalbi\inst{1}, S.~Colafrancesco\inst{4}, A. Cucchiara~\inst{5}, 
L.~Fuhrmann\inst{6}, H.A.~Krimm\inst{7}, N.~Marchili\inst{6}, 
E.~Massaro\inst{8}, M.~Perri\inst{1}, G.~Tagliaferri\inst{9}, 
G.~Tosti\inst{6}, A.~Tramacere\inst{8},
D.N.~Burrows\inst{5}, G.~Chincarini \inst{9}, A.~Falcone\inst{5}, 
N.~Gehrels\inst{7}, J.~Kennea\inst{5}, R. Sambruna\inst{7}\\
\institute{
            ASI Science Data Center, ASDC c/o ESRIN, 
            via G. Galilei 00044 Frascati, Italy. 
\and 
            Agenzia Spaziale Italiana, 
            Unit\`a Osservazione dell'Universo,             
\and
           UCL, Mullard Space Science Laboratory, Holmbury St. Mary, 
           Dorking, Surrey RH5 6NT, UK.
\and
           INAF - Osservatorio Astronomico di Roma 
           via Frascati 33, I-00040 Monteporzio, Italy. 
\and 
           Department of Astronomy and Astrophysics,  Pennsylvania  State 
           University, USA.  
\and
           Dipartimento di Fisica, Universit\'a di Perugia, Via A. Pascoli, 
           Perugia, Italy.
\and       NASA/Goddard Space Flight Center, Greenbelt, Maryland 20771, USA.   
\and       Dipartimento di Fisica, Universit\`a "La Sapienza",
            P.le A. Moro 2, I-00185 Roma, Italy. 
\and
           INAF - Osservatorio Astronomico di Brera 
            via Bianchi 46, 23807 Merate, Italy.         
 }}
\offprints{paolo.giommi@asi.it}
\date{Received ....; accepted ....}

\markboth{P. Giommi et al.:  \sw observations of the blazar 3C 454.3  
during a remarkably large optical and X-ray flare}
{P. Giommi et al.: \sw observations of the blazar 3C 454.3 during a 
remarkably large optical and X-ray flare}

\abstract{We present the results of a series of \sw and quasi simultaneous 
ground-based infra-red  observations of the blazar \3c carried out in 
April-May 2005 when the source was 10 to 30 times brighter than previously 
observed. 
We found \3c to be very bright and variable at all frequencies covered by 
our instrumentation.
The broad-band Spectral Energy Distribution (SED) shows the usual two-bump 
shape (in $\nu-\nu f(\nu)$  space) with the Infra-red, optical and UV data 
sampling the declining part of the synchrotron emission that, even during 
this extremely large outburst, had its maximum in the far-infrared. 
The X-ray spectral data from the XRT and BAT instruments are flat and due 
to inverse Compton emission. 
The remarkable SED observed implies that at the time of the \sw 
pointings \3c~ was one of the brightest objects in the extragalactic sky 
with a $\gamma$-ray emission similar or brighter than that of 3C~279 when 
observed in a high state by EGRET.  
Time variability in the optical-UV flux is very different from that in the 
X-ray data: while the first component varied by about a factor two within 
a single exposure, but remained approximately constant between different 
observations, the inverse Compton component did not vary on short time-scales 
but changed by more than a factor of 3 between observations separated by a 
few days. 
This different dynamical behaviour illustrates the need to collect simultaneous
multi-frequency data over a wide range of time-scales to fully constrain 
physical parameters in blazars.

}
\authorrunning{P. Giommi et al.}
\titlerunning{\sw observations of the blazar 3C 454.3
}

\maketitle

\section{Introduction}

\3c~ is a well known bright ($f_{(5GHz)}  \approx 10-15$Jy), moderately high 
redshift ($z=0.859 $) Flat Spectrum Radio Quasar (FSRQ) which shows all the 
typical hallmarks of the class of blazars: 
large intensity variations at all frequencies, high radio and optical 
polarization, superluminal motion and a Spectral Energy Distribution (SED) 
showing two broad peaks attributed to synchrotron and inverse Compton 
radiation. 
The synchrotron power peaks in the Infra-red band while Inverse Compton 
radiation starts at soft X-ray frequencies and peaks at MeV energies 
\citep{Gio02c, Blom95}.

Because of its brightness \3c~ has been extensively observed over the years 
in most energy bands, from radio \citep[e.g.][]{Bennett62}, through 
microwave \citep[WMAP][]{bennett03},  optical \citep[e.g.][]{Sandage66, 
Raiteri98}, X-ray \citep[e.g.][]{Worrall87,Tav02, Marshall05}, low energy 
$\gamma$-ray  \citep[COMPTEL,][]{Blom95, Zhang05} and high energy 
$\gamma$-ray \citep[EGRET,][]{Hartman93, Hartman99}.

\begin{figure}[t]
\includegraphics[width=6cm, angle=-90] {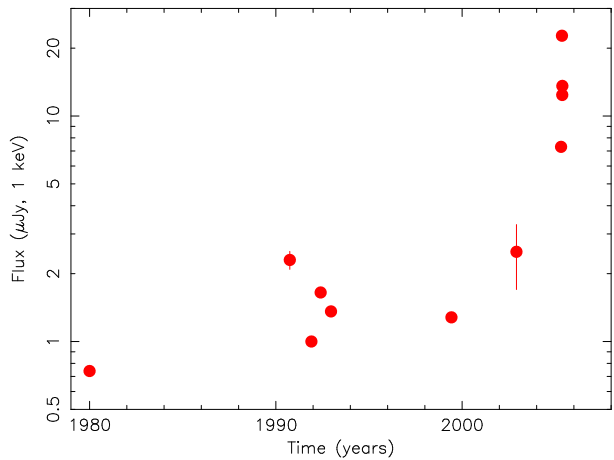}
\caption[]{
Long Term X-ray lightcurve of \3c~ at 1 keV, built using archival data from the {\it Einstein}, ROSAT, \sax, Chandra and \sw satellites. 
}
\label{lightcurve}
\end{figure}

In May 2005 \3c~ was reported to undergo a very strong optical flare with a 
remarkable flux increase of about four magnitudes compared to previous 
observations \citep{Balonek05a, Balonek05b}. 
During the same period the RXTE all-sky monitor recorded a flux of over 
10 mCrab, implying that \3c~ was extremely active also at X-ray frequencies 
where it had become one of the brightest extragalactic sources in the sky 
\citep{Remillard05}. 

The \sw satellite \citep{Gehrels04} pointed \3c~ on four occasions in 
April-May 2005, initially as part of an on-going project to study the X-ray 
properties of a sample of  blazars and then as a Target of Opportunity (ToO) 
following the announcement of the optical and X-ray outburst.  

Historically, \3c~ was observed on several occasions by a number of X-ray 
astronomy satellites, starting with the {\it Einstein} observatory in 1980 
up to the recent RXTE detection and \sw detailed observations.   
Fig. \ref{lightcurve} shows the long term X-ray lightcurve built using 
data from the {\it Einstein}, ROSAT,  \sax, Chandra and \sw satellites. 
The unusual and very large flare of spring 2005 is readily apparent.

Because of this exceptionally high state INTEGRAL observed \3c~ as a ToO in 
the hard-X/$\gamma$-ray band and detected it up to 200 keV \citep{Foschini05}.
The results of these observations are presented in \cite{pian06}. 
Optical and infra-red monitoring between May and August 2005 are reported 
by \cite{Fur06}.   

In this paper we report the results of the four \sw pointings and compare them 
with the results of quasi simultaneous optical and infrared observations 
performed with the ground-based REM telescope  \citep{Zerbi04}.  
We also report the hard X-ray light curve of \3c~ measured by the \sw BAT 
instrument when the blazar was not the target of the observation but was nevertheless 
within its very large field of view (1.4 sr, half-coded) and was bright enough to be detected. 
  
\section{\sw observations and data analysis}

\sw data have been collected using all three on-board experiments: 
the X-ray Telescope \citep[XRT,][]{Burrows05}, the UV and Optical Telescope 
\citep[UVOT,][]{Roming05} and the Burst Alert Telescope \citep[BAT,][]
{Barthelmy05}. 
These instruments, together with the observations from the ground-based 
REM Telescope provide a spectral coverage that ranges from the near 
infra-red to the hard X-rays. 

\subsection{XRT analysis}

The XRT observations were carried out using both the Photon Counting (PC) readout  mode, which provides maximum sensitivity but is affected by photon pile up 
for count-rates larger than $\approx 0.5 $ cts/s, and the Windowed Timing 
(WT) mode which does not provide full imaging capabilities but 
it does not suffer from pile-up up to count-rates of $\approx 200$ cts/s 
\citep[see][for details of the XRT observing modes]{Burrows05, Hill04}. 

The data were reduced using the {\it XRTDAS} software (v1.4.0) developed 
at the ASI Science Data Center (ASDC) and distributed within the HEAsoft 
6.0 package by the NASA High Energy Astrophysics Archive Research Center 
(HEASARC). 
We have selected photons with grades in the range 0-12 and used default 
screening parameters to produce level 2 cleaned event files.  

Since the source count rate (a few counts/s) was sufficiently large to cause 
photon pile-up in the central pixels of the PSF, the spectral data collected 
in the PC mode were extracted in an annular region with inner and outer radii 
of 6 and 20 pixels respectively \citep{Vaughan05}. 
The background was estimated in a nearby source-free circular region of 
50 pixels radius. 

Spectra in the data taken in WT mode were extracted in a $40\times20$ pixel 
rectangular region centered on the target. 
The background was estimated in a nearby $50\times20$ pixel source-free 
rectangular region.
In order to use $\chi^2$ statistics, spectra were rebinned to include at 
least 20 photons in  each energy channel. 
Due to current uncertainties in the XRT calibration at low energies photons below 
0.7 keV were excluded from the analysis.  

We used the XSPEC 11.3 spectral analysis package to fit the data to a simple 
power law spectrum.
Initially, we fixed the low energy absorption ($N_H$) to the Galactic 
value estimated from the 21 cm measurements of \citep[$6.5\times10^{20}$ 
cm$^{-2}$][]{Dick90}. 
The resulting spectra of all observations showed a systematic deviation
from the best fit law in the low energy part suggesting the possibility of
an additional absorption as already noticed by \citep{Tav02} or a spectral flattening below
 $\approx 1-2 $ keV. 
We then considered $N_H$ as a free parameter in the fitting and found a value
systematically higher than the Galactic one by a factor of 3.5 in
agreement with \cite{Tav02}.
The results are reported in Tab. \ref{tab.spectral_results} where column 
1 gives the observation date, column 2 gives the power law photon index, 
column 3 gives the flux in the 2-10 keV band, and column 4 gives the reduced 
$\chi^2$ and the number of degrees of freedom. 

A Chandra observation of \3c performed on 19-21 May 2005, shortly after our XRT observations,
 \citep{villata06} shows a similar X-ray flux level and also a low energy absorption in excess to that expected from Galactic $N_H$ , although lower than that observed by us.

\begin{table*}
\caption{Best fit power law spectral parameters for the \sw  and \sax ~observations of \3c. 
Numbers in parenthesis are statistical errors at the 90\% confidence level.}
\begin{center}
\begin{tabular}{llccccc}
\hline

 Date     &   Instrument &$\Gamma$ & N$_{\rm{H}}$&$F_{2-10keV}$ & $F_{15-150keV}$  & $\chi^2_r$/dof   \\
               &                       && cm$^{-2}$ &erg cm$^{-2}$s$^{-1}$ & erg cm$^{-2}$s$^{-1}$&  \\ 
               &                       & & ($10^{21}$) &($10^{-11}$) &($10^{-10}$)  & \\
               \\
\hline

 24 April 2005 & \sw XRT &  1.76 (0.07) & 2.3 (0.4) &$6.1\pm 0.3$  && 0.99/201  \\
 11 May 2005$^{*}$  &\sw XRT &  1.62 (0.08) &2.3 (0.4)& $16 \pm 1.0$ && 0.87/172   \\
 17 May 2005 & \sw XRT &1.79  (0.10) & 2.3 (0.6)&$8.5 \pm 0.5$  && 0.92/108 \\
 19 May 2005 & \sw XRT &1.71 (0.09) & 2.4 (0.6)&$10 \pm 1.0 $ && 0.83/122 \\
\hline
8 May 2005 & \sw BAT &1.7 (0.2) & --&  &5.9$\pm$1.0&0.68/57 \\
11 May 2005 & \sw BAT & -- & --&  &5.4$\pm$1.3& \\
17 May 2005 & \sw BAT & -- & --&  &$<$2.3& \\
19 May 2005 & \sw BAT & -- & --&  &3.3$\pm$0.6& \\
6 August 2005 & \sw BAT &1.8 (0.2) & --& & 1.7$\pm$0.3& 1.47/57 \\
\hline
5 June 2000 &  \sax -LECS+MECS &1.38  (0.06) &2.4(0.8)&  $1.1\pm 0.1$&& 1.12/58  \\

\hline

\multicolumn{5}{l} { $^{*}$ Windowed timing data}

\end{tabular}
\end{center}
\label{tab.spectral_results} 
\end{table*}

\subsection{UVOT observations}

The UVOT took data during all four \sw observations of \3c. 
The total exposure times in each of the UVOT filters is listed in 
Table~\ref{uvot_obs_summary}. 

During observations of April 24 and May 17 the UVOT obtained series of 
images in each of the lenticular filters (V, B, U, UVW1, UVM2, and UVW2). 
The UVOT was instead operated with the UV grism for the other two observations.

In the case of the lenticular filters, photometry was performed on the 
source using a standard tool (GAIA, the Starlink Graphical Astronomy and 
Image Analysis Tool; \citet{draper2004}). 
Counts were extracted from a 6\arcsec\, radius aperture (V, B and U filters) 
or a 10\arcsec\, radius aperture (UVW1, UVM2 and UVW2 filters); 
a larger radius was used for the UV filters due to the wider PSF in these 
filters. 
The count rate was corrected for coincidence loss (analogous to pile-up), 
and the background subtraction was performed using a background count rate 
obtained from a 20\arcsec\, radius region offset from the source. 
The count rates were then de-reddened using a value for $E(B-V)$ of 0.107 
mag \citep{schlegel1998} with the $A_{\lambda}/E(B-V)$ ratios given in 
Table~\ref{uvot_filters}, and converted to fluxes using the count rate to 
flux conversion factors also listed in Table~\ref{uvot_filters}. 
The resulting flux points for the V and UVW2 filters are plotted as a 
lightcurve in Fig. \ref{UVOT_lc} where we see that variability of up to 
a factor of 2 is present in the first observation (April 24).

  \begin{table}
      \caption[]{Total UVOT exposure time in seconds in the V, B, U, UVW1, 
UVM2, UVW2 filters or UV grism}
         \label{uvot_obs_summary}
         \scriptsize
         \begin{tabular}{lccccccc}
            \hline
            \noalign{\smallskip}
            Date &  Exp. & Exp & Exp & Exp & Exp & Exp & Exp \\
            &  V & B & U & UVW1 &UVM2 &UVW2 &Grism \\       
            \noalign{\smallskip}
            \hline
            \noalign{\smallskip}
24 Apr & 947 & 675 & 948 & 1892 & 2843 & 3926 & 0 \\
11 May & 0 & 0 & 0 & 0 & 0 & 0 & 1941 \\
17 May & 400 & 306 & 400 & 791 & 1197 & 1598 \\
19 May & 0 & 0 & 0 & 0 & 0 & 0 & 4964 \\
            \noalign{\smallskip}
            \hline
         \end{tabular}
  \end{table}

  \begin{table}
    
      \caption[]{Effective wavelengths, extinction ratios, and counts-to-flux conversion factors for the UVOT filters}
         \label{uvot_filters}
         \scriptsize
         \begin{tabular}{lccc}
            \hline
            \noalign{\smallskip}
Filter & $\lambda_{eff}^{\mathrm{a}}$ & $A_{\lambda_{eff}}/E(B-V)^{\mathrm{b}}$ & $F_{conv}^{\mathrm{c}}$  \\ 
            \noalign{\smallskip}
            \hline
            \noalign{\smallskip}
$V$    & 5460 & 3.2 & (2.2 $\pm$ 0.1) $\times 10^{-16}$   \\
$B$    & 4350 & 4.2 & (1.3 $\pm$ 0.1) $\times 10^{-16}$   \\
$U$    & 3450 & 5.2 & (1.5 $\pm$ 0.06) $\times 10^{-16}$   \\
$UVW1$ & 2600 & 6.7 & (3.5 $\pm$ 0.07) $\times 10^{-16}$   \\
$UVM2$ & 2200 & 9.7 & (7 $\pm$ 1) $\times 10^{-16}$   \\
$UVW2$ & 1930 & 8.3 & (6.0 $\pm$ 0.4) $\times 10^{-16}$   \\
            \noalign{\smallskip}
            \hline
         \end{tabular}
\begin{list}{}{}
\item[$^{\mathrm{a}}$] Effective wavelength of the filter in \AA. 
\item[$^{\mathrm{b}}$] Ratio of $A_{\lambda_{eff}}$, the extinction in 
magnitudes at the effective wavelength of the filter $\lambda_{eff}$, to 
$E(B-V)$. For the UV filters (UVW1, UVM2 and UVW2) $A_{\lambda_{eff}}/E(B-V)$ was 
obtained from Fig. 1 in \citet{seaton1979}, and for the optical (V, B and U) 
filters the values are derived from the interstellar reddening curve given by  \citet{zombeck1990}.
\item[$^{\mathrm{c}}$] Conversion factor from rate (Counts~s$^{-1}$) to flux 
(Erg~cm$^{-2}$s$^{-1}$\AA$^{-1}$) from UVOT in-orbit calibration as of 30 June
2005 (T.S. Poole, priv. comm.).

\end{list}
\end{table}

\subsection{BAT data}
Thanks to its very large field of view (1.4 sr, half-coded), the BAT 
instrument was able to monitor 3C 454.3 even when it was not the target of the \sw observation. 
The source was in the BAT field of view on most days starting in early April 2005 and 
continuing beyond the XRT and UVOT observations (see BAT light curve shown 
in Fig. \ref{bat_lc}).  
The source showed considerable variation, with significant detections in 
at least twenty observations including two of the pointed observations 
(May 11 and May 19).
The BAT lightcurve was accumulated for all observations in which 3C 454.3 
was close enough to the field centre that it illuminated at least 10\% of 
the BAT detectors. 
This corresponds to ~38 degrees from the centre in the short direction and 
~55 degrees in the long direction of the roughly rectangular BAT field of 
view. 
The count rate for each orbit was determined from the sky images by 
subtracting an annular background and fitting the BAT point spread function 
(PSF). 
The orbital count rates were combined to produce daily averages. 
For each day on which the signal to noise ratio was at least 4.0, spectra 
were produced in the following analysis chain using the Swift-BAT software. 
The BAT survey detector plane histograms (80 energy channels, accumulated 
over a single spacecraft pointing) were first rebinned in energy to correct 
a small residual inaccuracy arising from the way the histograms are created 
on-board the instrument. 
Secondly, noisy detectors with a count rate more than ~twice the mean rate 
were removed from the analysis data set. 
Thirdly, the background and other bright sources (such as Cygnus X-1) in the 
field of view were removed by "cleaning" in which a geometric model of 
the background plus extraneous sources (coded by the BAT mask) are fit to the 
detector array and then subtracted. 
Then all detectors lying on the projection of bright sources through the 
edge of the coded aperture were removed from the analysis data set. 
This step is necessary because the geometry and absorption of gaps in the 
coded aperture and graded-Z shield near the edges are not well known 
and hence cannot be well-modeled in the response function.
Finally counts spectra are created using BATBINEVT by convolving the 
detector plane histograms with a weight map giving the fraction of each 
BAT detector exposed to the source through the empty cells in the BAT coded 
aperture.
The spectra for individual pointings in an observation were combined using 
the MATHPHA tool.
The response function was generated for each observation using the \sw 
BATDRMGEN routine. 
It correctly accounts for disabled and ignored detectors and geometric effects
when 3C 454.3 is not in the centre of the BAT field of view.
The spectra were then fit to a simple power law model using the XSPEC 11.3 
analysis package.
Only during two observations (8 May and 6 August 2005) the data were good 
enough to allow us to estimate the power law spectral index with  a statistical 
error less than 0.25. 
On these occasions the best fit spectral indices were $\Gamma=1.69\pm 0.19$ 
and $\Gamma=1.81\pm 0.22$, the fluxes in the 15-150 keV were $5.9\times 10^{-10}$
\ergs and $1.7\times 10^{-10}$ \ergs, respectively (see also Tab.\ref{tab.spectral_results}), 
in good agreement with the extrapolations of XRT spectra.
For completeness we also report in Tab.\ref{tab.spectral_results} BAT fluxes simultaneous 
with XRT observations even when no spectral fitting was possible. In these cases fluxes 
were calculated converting the BAT count-rate into 15-150 keV flux assuming a spectral 
index of 1.7 as measured during the observation of 8 May 2005.

\section{A reanalysis of the \sax~ observation}

\3c~ was observed  by \sax~ in June 2000 when the source was about 20 times 
fainter than during the \sw observation of May 11. 
Despite the relatively low flux the LECS, MECS and PDS instruments provided 
a good quality X-ray spectrum between 0.1 to over 100 keV \citep[see][for a 
description of the instrumentation]{Boella97a}.  
These 
broad-band 
X-ray data (first reported by \citet{Tav02} and subsequently 
by \citet{Gio02c}) are useful to compare the SED of this source in different 
brightness states. 

\begin{table}[t]

\caption{Summary of REM~ observations}
\label{tab1}
\begin{center}
\begin{tabular}{lcrr}
\hline
 Date     & Start UT & Filter  & magnitude  \\
          &          &       &    \\
\hline

11 May   & 21:41 & $R$  & 12.73 (0.04)  \\
11 May   & 21:45 & $I$  & 12.04 (0.07)  \\
11 May   & 21:50 & $V$  & 13.28 (0.08)  \\
17 May   & 21:43 & $H$  & 9.74 (0.04)  \\
18 May   & 09:41 & $R$  & 12.56 (0.05)  \\
18 May   & 09:46 & $I$  & 11.84 (0.05)  \\
18 May   & 09:51 & $V$  & 13.08 (0.07)  \\
19 May   & 09:12 & $R$  & 12.22 (0.07)  \\
19 May   & 09:17 & $I$  & 11.58 (0.07)   \\
19 May   & 09:22 & $V$  & 12.72 (0.07)  \\
19 May   & 21:12 & $H$  & 9.94 (0.03)  \\
\hline
\multicolumn{4}{c} { }
\end{tabular}
\end{center}
\label{tab.REM}

\end{table}

We present here a re-analysis of the {\it Beppo}SAX data considering a 
single power law model with free low energy absorption.
We considered LECS, MECS and PDS data. 
Events for spectral analysis were selected in circular regions of 4 and 3 
arcminutes and in the energy bands 0.5--2.0 keV and 2.0--10.0 keV for the 
LECS and MECS, respectively. 
PDS data in the 15-100 keV range were used. 
Background spectra were taken from the blank field archive at the ASI Science 
Data Center.

In lowest line of Table \ref{tab.spectral_results} we reported the best fit 
parameters of the {\it Beppo}SAX observation. 
As can be seen, the model describes well the 0.5-100 keV spectrum and the 
best fit column density is significantly higher than the Galactic value, 
confirming the early report by \citet{Tav02} and in good agreement 
with the XRT observations of 2005 (See tab \ref{tab.spectral_results}).

\section{REM observations}

The Rapid Eye Mount \citep[REM,][]{Zerbi04} is a robotic telescope located 
at the ESO Cerro La Silla observatory (Chile).
The REM  telescope has a Ritchey-Chretien configuration with a 60 cm 
f/2.2 primary and an overall f/8 focal ratio in a fast moving alt-azimuth 
mount providing two stable Nasmyth focal stations. 
At one of the two foci the telescope simultaneously feeds, by means of 
a dichroic, two cameras:
REMIR for the NIR \citep[see][]{Conconi04}, and ROSS \citep[see][]{Tosti04} 
for the optical.
Both cameras have a 10 x 10 arcmin field of view and imaging capabilities 
with the usual NIR ($z'$, $J$, $H$ and $K$) and Johnson-Cousins $VRI$ filters.
Moreover, low resolution slit-less spectroscopy is also possible via an 
Amici prism.

All raw optical/NIR frames obtained by REM were corrected for dark, bias 
and flat field. Instrumental magnitudes were obtained via aperture photometry 
using DAOPHOT \citep{Stetson88} and Sextractor \citep{Bertin96}. 
Calibration of the optical source magnitude was obtained by differential 
photometry with respect to the comparison stars sequence reported by 
\cite{Fiorucci98} and \cite{Raiteri98}. 
For the NIR calibration we used the comparison sequence reported by 
\cite{Gonzalez01}.
 
The observed fluxes in different filters are reported in Tab. \ref{tab.REM} 
where columns 1 and 2 give the data and time of the observation, column 3 
gives the filter and column 5 gives the magnitude with error in parenthesis.

\begin{figure}[t] 
\includegraphics[width=8cm, angle=0] {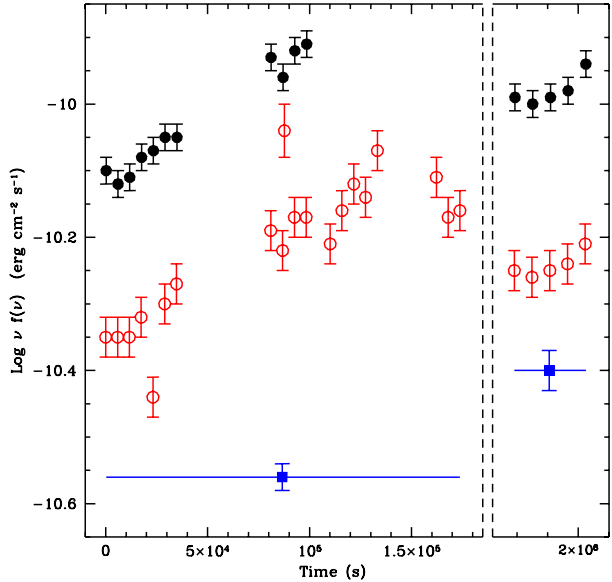}
\caption[]{UVOT light curve of \3c~ in the $V$ (filled circles) and $UVW2$ 
(open circles) filters during the observations of April 24 (left side) and May 17 (right side).  
The time axis is in units of seconds since the beginning of the first UVOT observation (24-April-2005 19:25:27 UT). Similar variability is present in all other UVOT filters. For comparison we also report as filled squares the simultaneous (constant during single exposures) 2-10 keV intensity level.}
\label{UVOT_lc}
\end{figure}

\begin{figure}[t] 

\includegraphics[width=6.3cm, angle=-90] {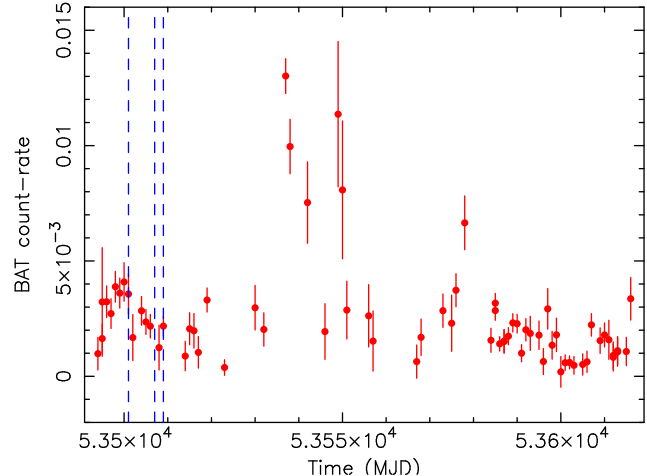}

\caption[]{BAT light-curve of 3C 454.3 between 25 April and 15 August 2005. The dates of the \sw pointed observations of May 11, 17 and 19 are indicated by the dashed vertical lines.}
\label{bat_lc}
\end{figure}

\section{Discussion and conclusion}

We have reported the results of several \sw and REM observations of the 
blazar \3c~ taken in April-May 2005 when the source undewent an extremely 
large optical and X-ray outburst. 

The source was found to be much brighter than in the past over the entire 
energy range covered by our instrumentation, from the near infra-red to the 
hard X-rays. 

The  maximum observed X-ray flux was a factor $\sim$10 larger than that 
measured during the ROSAT All Sky Survey and a factor $\sim$30 higher than 
during an early {\it Einstein} observation.  
Comparison with the 2002 Chandra observation is somewhat uncertain due 
to pile-up problems in the ACIS instrument \citep{Marshall05}. 

The time variability behaviour of the IR-optical-UV flux (thought to be 
due to synchrotron emission) and of the X-ray flux (attributed to the 
inverse Compton component) were quite different.
While the X-ray flux was stable within single exposures, but varied by over 
a factor 3 between different \sw observations, the optical/UV flux varied 
by nearly a factor 2 during the observation of April 24 (see Fig. 
\ref{UVOT_lc})  but the average flux varied much less than the X-ray flux 
between different observations. 

\begin{figure*}[t] 
\begin{center}
\includegraphics[width=11.cm, angle=-90] {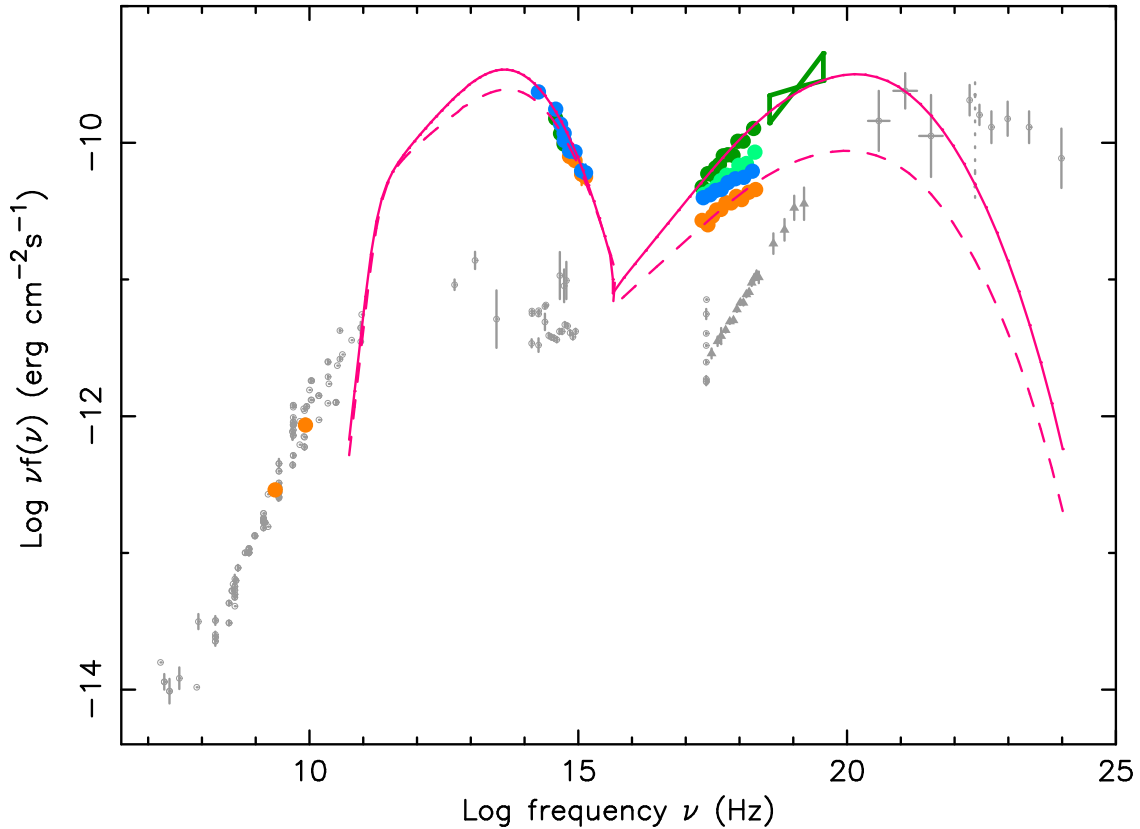}

\caption[]{The observed Spectral Energy Distribution of \3c. \sw UVOT, XRT and BAT data 
are plotted together with quasi-simultaneous REM data (large, filled circles) 
and with non-simultaneous multi-frequency data (smaller symbols). 
BAT spectral results, shown as a bow-tie spectrum, are from the observation 
of 8 May 2005 when the source intensity was very large. The solid and dashed  lines represent the predictions of simple one-zone SSC models that fit the optical and X-ray data reasonably well. Detailed modelling, however, requires a more complex approach that takes into account the different dynamical time-scales at different frequencies. See text for details.
}
\end{center}
\label{SED}
\end{figure*}

The X-ray spectral slope was found to be flat during all four \sw 
observations with some indication of spectral flattening with increasing 
intensity. 
Best fits to simple power law models give photon spectral indices of 
$\Gamma \approx 1.6-1.8$, strongly indicating that the inverse Compton 
component was responsible for the X-ray emission also in this very bright 
state (see Table \ref{tab.spectral_results}). 
Similar results have been obtained with INTEGRAL observations perfomed on 
15-18 May 2005 \citep{pian06}.

We have combined our XRT, UVOT and REM measurements with non-simultaneous 
multifrequency data taken form NED and the ASI Science Data Center to build 
the SED shown in Fig. \ref{SED}. 
From this figure we see that the infra-red, optical and UV data points form 
a steep spectrum implying that the peak of the synchrotron component is 
located at lower energies than the near infrared even during the flare 
\citep[see also ][]{Fur06}.
\3c~ was detected by COMPTEL and EGRET in the $\gamma$-ray band a few years 
ago when the source was much fainter in the optical band than during our 
observations \citep[e.g.][]{Raiteri98}. 

The general shape of the SED clearly follows the usual two-bump synchrotron/inverse Compton scenario. 
We computed the expected SED from one-zone synchrotron self-Compton models from 
a power law distribution of relativistic electrons with energy up to $\gamma = 10^3$ followed 
by a log-parabolic distribution up to a Lorentz factor of  $\gamma_{max}=3\times 10^4$  \citep{Massaro06a}. We obtained reasonably good representations of the data for both the high and low intensity states (see Fig. 4)
 assuming magnetic field values of $B\simeq 0.3$ and $0.45~$Gauss, beaming factor $\delta \simeq 20$, 
 electrons power law  slopes $s = -2$ and $- 2.3 $ and curvature parameter $r= 1.8$ and $1.7$  \citep[see][for details]{Massaro06a}.  However, the widely different behaviour of the apparently constant optical component and the variable X-ray flux undoubtly indicate the presence of widely different dynamical time-scales in different energy bands. Moreover, the optical light-curve reported by  \citet{villata06} showing large variability, both on relatively long (several weeks) and short timescales  (see also Fig. \ref{UVOT_lc}), implies an even more complex scenario. 
A complete understanding of the physical processes powering flares such as that observed on 
May 2005 require the measurement of the dynamical time-scales at different frequencies.
Detailed modelling of the data presented here is beyond the scope of the present work.

Despite \3c~ was about 20 times brighter than when observed with \sax~ the 
X-ray spectral shape was significantly harder when the source was fainter. 
This finding contrasts with what is usually seen during flares in blazars
and suggests that it could be due the occurrence of a softer IC component.
In fact, as apparent from Fig. 4
we cannot exclude that the peak of the SED, 
which can be estimated around 10 MeV in the faint state moved down 
to $\sim$1 MeV during the 2005 flare. 
Unfortunately, the lack of  $\gamma$-ray data makes it impossible to reach a firm conclusion
on this point that is crucial to model the physical conditions of \3c.

On the other hand, if the $\gamma$-ray flux scaled with the optical flux \3c~ might have been the brightest 
AGN in the May 2005 $\gamma$-ray sky with a flux in excess of $10^{-6}$ \phcm2 (E $>$ 100 MeV), probably brighter than 3C~279 when observed by EGRET in a high state.
Differently, a spectral distribution of the flaring component with a more
pronounced curvature would imply a $\gamma$-ray flux of the same order or 
even lower than that observed by EGRET.
Blazars undergoing large outbursts like the one described in this paper 
are obvious targets for the upcoming $\gamma$-ray observatories AGILE 
and GLAST. 
Observations with these satellites, combined with well planned multi-frequency
monitoring campaigns, will provide great new opportunities to significantly 
improve our understanding of the physical processes powering blazars.
  
\begin{acknowledgements}

The authors acknowledge financial support from the Italian Space Agency (ASI) 
through grant I/R/039/04 and through funding of the ASDC.
The UCL-MSSL authors acknowledge support of PPARC. 
This work is partly based and on data taken from the NASA/IPAC Extragalactic 
Database (NED) and from the ASI Science Data Center (ASDC). We thank the anonymous referee
for useful suggestions.
\end{acknowledgements}

\begin{thebibliography}{}
\expandafter\ifx\csname natexlab\endcsname\relax\def\natexlab#1{#1}\fi

\bibitem[{Balonek(2005{\natexlab{a}})}]{Balonek05a}
Balonek, T. 2005{\natexlab{a}}, vsnet -alert 8383

\bibitem[{Balonek(2005{\natexlab{b}})}]{Balonek05b}
Balonek, T. 2005{\natexlab{b}}, vsnet -alert 8405

\bibitem[{Barthelmy {et~al.}(2005)Barthelmy, Barbier, Cummings, Fenimore,
  Gehrels, \& et~al.}]{Barthelmy05}
Barthelmy, S., Barbier, L.~M., Cummings, J., {et~al.} 2005, SSRv., 120, 95

\bibitem[{Bennett(1962)}]{Bennett62}
Bennett, A.~S. 1962, MNRAS, 68, 163

\bibitem[{Bennett {et~al.}(2003)Bennett, Hill, Hinshaw, Nolta, Odegard, Page,
  Spergel, Weiland, Wright, Halpern, \& coauthors}]{bennett03}
Bennett, C.~L., Hill, R.~S., Hinshaw, G., {et~al.} 2003, ApJS, 148, 97

\bibitem[{Bertin \& Arnouts(1996)}]{Bertin96}
Bertin, E. \& Arnouts, S. 1996, A\&AS, 117, 393

\bibitem[{Blom {et~al.}(1995)Blom, Bloemen, Bennett, Collmar, Hermsen,
  McConnell, Schoenfelder, Stacy, Steinle, Strong, \& Winkler}]{Blom95}
Blom, J., Bloemen, H., Bennett, K., {et~al.} 1995, A\&A, 295, 330

\bibitem[{Boella {et~al.}(1997)Boella, Butler, Perola, Piro, Scarsi, \&
  Bleeker}]{Boella97a}
Boella, G., Butler, R.~C., Perola, G.~C., {et~al.} 1997, A\&AS, 122, 229

\bibitem[{Burrows {et~al.}(2005)Burrows, Hill, Nousek, Kennea, Wells, Osborne,
  Abbey, Beardmore, Mukerjee, \& et~al.}]{Burrows05}
Burrows, D., Hill, J.~E., Nousek, J.~A., {et~al.} 2005, SSRv., 120, 165

\bibitem[{Conconi {et~al.}(2004)Conconi, Cunniffe, D'Alessio, \&
  et~al.}]{Conconi04}
Conconi, P., Cunniffe, R., D'Alessio, G., \& et~al. 2004, SPIE, 5492, 1602

\bibitem[{Dickey \& Lockman(1990)}]{Dick90}
Dickey, J. \& Lockman, F. 1990, ARA\&A, 28, 215

\bibitem[{Draper {et~al.}(2004)Draper, Gray, \& Berry}]{draper2004}
Draper, P.~W., Gray, N., \& Berry, D.~S. 2004, starlink User Note 214.15

\bibitem[{Fiorucci {et~al.}(1998)Fiorucci, Tosti, \& Rizzi}]{Fiorucci98}
Fiorucci, M., Tosti, G., \& Rizzi, N. 1998, PASP, 110, 105

\bibitem[{Foschini {et~al.}(2005)Foschini, Di~Cocco, Malaguti, Pian,
  Tagliaferri, Ghisellini, Maraschi, Giommi, Gehrels, Walter, Eckert, \&
  on~behalf of~a Large~Collaboration}]{Foschini05}
Foschini, L., Di~Cocco, G., Malaguti, G., {et~al.} 2005, aTel, 497

\bibitem[{Fuhrmann {et~al.}(2006)Fuhrmann, Cucchiara, Marchili, Tosti, \& et.
  al.}]{Fur06}
Fuhrmann, L., Cucchiara, A., Marchili, N., Tosti, G., \& et. al. 2006, A\&A,
  445, L1

\bibitem[{Gehrels {et~al.}(2004)Gehrels, Chincarini, Giommi, Mason, Nousek,
  Wells, White, Barthelmy, Burrows, Cominsky, \& coauthors}]{Gehrels04}
Gehrels, N., Chincarini, G., Giommi, P., {et~al.} 2004, ApJ, 611, 1005

\bibitem[{Giommi {et~al.}(2002)Giommi, Capalbi, Fiocchi, Memola, Perri,
  Piranomonte, Rebecchi, \& Massaro}]{Gio02c}
Giommi, P., Capalbi, M., Fiocchi, M., {et~al.} 2002, in Blazar Astrophysics
  with {\it BeppoSAX} and Other Observatories, ed. P.~Giommi, E.~Massaro, \&
  G.~Palumbo, 63

\bibitem[{Gonzalez-Perez {et~al.}(2001)Gonzalez-Perez, Kidger, Martin-Luis, \&
  et~al.}]{Gonzalez01}
Gonzalez-Perez, J.~N., Kidger, M., Martin-Luis, F., \& et~al. 2001, AJ, 122,
  2055

\bibitem[{Hartman {et~al.}(1999)Hartman, Bertsch, Bloom, Chen, Deines-Jones,
  Esposito, Fichtel, Friedlander, Hunter, McDonald, Sreekumar, Thompson, Jones,
  Lin, Michelson, Nolan, Tompkins, Kanbach, Mayer-Hasselwander, Mücke, Pohl,
  Reimer, Kniffen, Schneid, von Montigny, Mukherjee, \& Dingus}]{Hartman99}
Hartman, R.~C., Bertsch, D.~L., Bloom, S.~D., {et~al.} 1999, APJS, 123, 79

\bibitem[{Hartman {et~al.}(1993)Hartman, Bertsch, Dingus, Fichtel, Hunter,
  Kanbach, Kniffen, Lin, Mattox, Mayer-Hasselwander, Michelson, von Montigny,
  Nolan, Piner, Schneid, Sreekumar, \& Thompson}]{Hartman93}
Hartman, R.~C., Bertsch, D.~L., Dingus, B.~L., {et~al.} 1993, ApJ, 407, L41

\bibitem[{Hill(2004)}]{Hill04}
Hill, J.E., e.~a. 2004, SPIE, 5165, 217

\bibitem[{Marshall {et~al.}(2005)Marshall, Schwartz, Lovell, Murphy, Worrall,
  Birkinshaw, Gelbord, Perlman, \& Jauncey}]{Marshall05}
Marshall, H.~L., Schwartz, D.~A., Lovell, J. E.~J., {et~al.} 2005, ApJS, 156,
  13

\bibitem[{Massaro {et~al.}(2006)Massaro, Tramacere, Perri, Giommi, \&
  Tosti}]{Massaro06a}
Massaro, E., Tramacere, A., Perri, M., Giommi, P., \& Tosti, G. 2006, A\&A,
  448, 861

\bibitem[{Pian {et~al.}(2006)Pian, Foschini, Beckmann, Soldi, T\~urler, \&
  et~al.}]{pian06}
Pian, E., Foschini, A., Beckmann, V., {et~al.} 2006, A\&A, 449, L21

\bibitem[{Raiteri {et~al.}(1998)Raiteri, Ghisellini, Villata, de~Francesco,
  Lanteri, Chiaberge, Peila, \& Antico}]{Raiteri98}
Raiteri, C.~M., Ghisellini, G., Villata, M., {et~al.} 1998, A\&AS, 127, 445

\bibitem[{Remillard(2005)}]{Remillard05}
Remillard, R. 2005, aTel, 484

\bibitem[{Roming {et~al.}(2005)Roming, Kennedy, Mason, Nousek, Ahr, \& et.
  al.}]{Roming05}
Roming, P. W.~A., Kennedy, T.~E., Mason, K.~O., {et~al.} 2005, SSRv, 120, 143

\bibitem[{Sandage(1966)}]{Sandage66}
Sandage, A. 1966, ApJ, 144, 1234

\bibitem[{Schlegel {et~al.}(1998)Schlegel, Finkbeiner, \& Davis}]{schlegel1998}
Schlegel, D.~J., Finkbeiner, D.~P., \& Davis, M. 1998, ApJ, 500, 525

\bibitem[{Seaton(1979)}]{seaton1979}
Seaton, M.~J. 1979, MNRAS, 187, 73

\bibitem[{Stetson(1988)}]{Stetson88}
Stetson, P. 1988, PASP, 99, 191

\bibitem[{Tavecchio {et~al.}(2002)Tavecchio, Maraschi, Ghisellini, Celotti,
  Comastri, Fossati, Grandi, Pian, \& Tagliaferri}]{Tav02}
Tavecchio, F., Maraschi, L., Ghisellini, G., {et~al.} 2002, ApJ, 575, 137

\bibitem[{Tosti {et~al.}(2004)Tosti, Bagaglia, Campeggi, \& et~al.}]{Tosti04}
Tosti, G., Bagaglia, M., Campeggi, C., \& et~al. 2004, SPIE, 5492, 689

\bibitem[{Vaughan {et~al.}(2005)Vaughan, Goad, Beardmore, O'Brien, Osborne, \&
  several~other authors}]{Vaughan05}
Vaughan, S., Goad, M.~R., Beardmore, A.~P., {et~al.} 2005, submitted to ApJ

\bibitem[{Villata {et~al.}(2006)Villata, Raiteri, Balonek, \& the
  WEBT~collaboration}]{villata06}
Villata, M., Raiteri, C., Balonek, T., \& the WEBT~collaboration. 2006, A\&A,
  in press, astro-ph/0603386

\bibitem[{Worrall {et~al.}(1987)Worrall, Tananbaum, Giommi, \&
  Zamorani}]{Worrall87}
Worrall, D.~M., Tananbaum, H., Giommi, P., \& Zamorani, G. 1987, ApJ, 313, 596

\bibitem[{Zerbi {et~al.}(2004)Zerbi, Chincarini, Ghisellini, \&
  et~al.}]{Zerbi04}
Zerbi, F., Chincarini, G., Ghisellini, G., \& et~al. 2004, SPIE, 5492, 1590

\bibitem[{Zhang {et~al.}(2005)Zhang, Collmar, \& Scho\"nfelder}]{Zhang05}
Zhang, S., Collmar, W., \& Scho\"nfelder. 2005, A\&A, 444, 767

\bibitem[{Zombeck(1990)}]{zombeck1990}
Zombeck, M.~V. 1990, Handbook of Astronomy and Astrophysics (Cambridge, UK:
  Second edition, Cambridge University Press)

\end{thebibliography}

\end{document}